\documentclass{crm-eca}

\usepackage[cmtip,all]{xy}
\usepackage{hyperref}
\usepackage{tipa}

\usepackage{type-theory}  

\newtheorem{theorem}{Theorem}

\theoremstyle{definition}
\newtheorem{definition}[theorem]{Definition}

\theoremstyle{remark}

\begin{document}

\papertitle{Classical field theory \\ via Cohesive homotopy types}


\paperauthor{Urs Schreiber}
\paperaddress{University Nijmegen} 
\paperemail{\href{http://ncatlab.org/nlab/show/Urs+Schreiber}{ncatlab.org/nlab/show/Urs+Schreiber}}




\makepapertitle

In the year 1900, at the {International Congress of Mathematics} in Paris,
David Hilbert stated his famous list of 23 central open questions of mathematics
\cite{Hi1900}. Among them, the sixth problem  (see \cite{Corry04} for a review)
is arguably the one that Hilbert himself regarded as the
most valuable:
``From all the problems in the list, the sixth is the only one that continually engaged
[Hilbert's] efforts over a very long period, at least between 1894 and 1932.''
\cite{Corry06}.
Hilbert stated the problem as follows

\vspace{4pt}

\noindent{\bf Hilbert's mathematical problem 6.}
{\it To treat by means of {axioms}, those {physical sciences} in
	which mathematics plays an important part}.

\vspace{4pt}

Since then, various aspects of physics have been given a mathematical formulation.
The following table, necessarily incomplete, gives a broad idea of central
concepts in theoretical physics and the mathematics
that captures them.

\vspace{.5cm}

\begin{tabular}{|l|ll|}
    \hline
    & {\bf physics} & {\bf maths}
    \\
	\hline\hline
    & {\it prequantum physics} & {\it differential geometry}
    \\
	\hline\hline
    18xx-19xx & {mechanics} & {symplectic geometry}
	\\
	1910s & {gravity} &
	{Riemannian geometry}
	\\
	1950s & {gauge theory} &
	{Chern-Weil theory}
	\\
	2000s & {higher gauge theory} & {differential cohomology}
	\\
	& &
	\\
	\hline \hline
    & {\it quantum physics} & {\it noncommutative algebra}
    \\
	\hline\hline
	1920s & {quantum mechanics} & operator algebra
	\\
	1960s & {local observables}  & co-sheaf theory
    \\
    1990s-2000s & {local field theory}
    & {$(\infty,n)$-category theory}
	\\
	\hline
\end{tabular}

\vspace{.5cm}

\noindent These are traditional solutions to aspects of Hilbert's sixth  problem.
Two points are noteworthy: on the one hand the items in the list are
crown jewels of mathematics; on the other hand their appearance is somewhat
unconnected and remains piecemeal.

\medskip

Towards the end of the 20th century, William Lawvere, the founder
of categorical logic and of categorical algebra, aimed for a more
encompassing answer that rests the axiomatization of physics on a decent
unified foundation. He suggested to
\begin{enumerate}
  \item rest the foundations of mathematics itself in topos theory
     \cite{Lawvere65};
  \item build the foundations of physics \emph{synthetically} inside topos theory by
  \begin{enumerate}
    \item
  imposing properties on a topos which ensure that the
  objects  have the
  structure of \emph{differential geometric spaces}
  \cite{Lawvere98};
  \item formalizing classical mechanics on this basis by
        universal constructions
        \\
  (``Categorical dynamics'' \cite{Lawvere67},
   ``Toposes of laws of motion'' \cite{LawvereSynth}).
\end{enumerate}
\end{enumerate}
\noindent While this is a grandiose plan, we have to note that it falls short in two respects:
\begin{enumerate}
  \item Modern mathematics prefers to refine its foundations from
  topos theory to \emph{higher topos theory}
  \cite{Lurie} viz. \emph{homotopy type theory} \cite{HoTT}.
  \item Modern physics needs to refine classical mechanics to
  \emph{quantum mechanics} and \emph{quantum field theory}
  at small length/high energy scales \cite{Feynman85, SatiSchreiber}.
\end{enumerate}
Concerning the first point, notice that indeed, as conjectured in \cite{Joyal} and proven by \cite{CisinskiShulman}:

\noindent{\it Homotopy type theory is the internal language of
locally Cartesian closed $\infty$-categories $\mathcal{C}$.}
\noindent Moreover \cite{Shulman}: {\it The univalence axiom encodes the presence of the small object classifier
in locally cartesian closed $\infty$-categories $\mathcal{C}$ which are in fact $\infty$-toposes $\mathbf{H}$.}

\vspace{8pt}

Therefore our task is to: refine Lawvere's synthetic approach on Hilbert's sixth
problem from
classical physics formalized in synthetic differential geometry
axiomatized in topos theory to
high energy physics formalized in higher differential geometry
axiomatized in higher topos theory.
Specifically, the task is to add to (univalent) homotopy type theory axioms that
make the homotopy types have the interpretation of differential \emph{geometric homotopy types}
in a way that admits a formalization of high energy physics.

The canonical way to add such \emph{modalities} on type theories is to add \emph{modal operators}
which in homotopy type theory are \emph{homotopy modalities} \cite{ShulmanMod}.
The $\infty$-categorical semantics of a homotopy modality is an idemponent $\infty$-(co-)monad as in \cite{Lurie}.
For these it is clear what an \emph{adjoint pair} is. We say:
\begin{definition}[\cite{ScSh}]
  \label{cohesion}
  \emph{Cohesive homotopy type theory} is univalent homotopy type theory
  equipped with an adjoint triple of homotopy (co-)modalities
  $
    \mbox{$\int$} \hspace{3pt} \dashv \hspace{3pt} \flat  \hspace{3pt}\dashv \hspace{3pt} \sharp
  $,
  to be called:
  $
    \mbox{\emph{shape modality}} \hspace{3pt} \dashv \hspace{3pt} \mbox{\emph{flat co-modality}}\hspace{3pt} \dashv  \hspace{3pt}\mbox{\emph{sharp modality}}
  $,
  such that there is a canonical equivalence of the $\flat$-modal types
  with the $\sharp$-modal types, and such that $\int$ preserves finite product types.
\end{definition}
This has been formalized in HoTT-Coq by Mike Shulman, see \cite{ScSh} for details.

With hindsight one finds that this modal type theory is essentially what Lawvere
was envisioning in \cite{Como}, where it is referred to as encoding
``being and becoming'', and later more formally in \cite{Lawvere94, Lawvere}, where it is referred to
as encoding ``cohesion''.

While def. \ref{cohesion} may look simple, its consequences are rich.
In \cite{dcct} we show how cohesive homotopy type theory synthetically captures not just
differential geometry, but the theory of (generalized) \emph{differential cohomology}
(e.g. \cite{BunkeDifferentialCohomology}). This is the cohomology theory in which physical gauge fields
(such as the field of electromagnetism) are cocycles. We show in \cite{dcct} that cohesion
implies the existence of geometric homotopy types $\mathbf{Phases}$ such that
\begin{enumerate}
\item the
dependent homotopy types over $\mathbf{Phases}$ are \emph{prequantized covariant phase spaces}
of physical field theories;
\item correspondences between these dependent types are
\emph{spaces of trajectories equipped with local action functionals};
\item
group actions on such dependent types encode the
Hamilton-de Donder-Weyl equations of motion of local covariant field theory;
\item the ``motivic'' linearization of these relations over suitable
stable homotopy types yields the corresponding quantum field theories.
\end{enumerate}
An exposition of what all this means is in section 1.2 of \cite{dcct}. See
\cite{Nuiten} for details on the last point. See \cite{syntheticQFT} for a general overview.

Specifically, cohesive homotopy type theory has semantics in the $\infty$-topos $\mathbf{H}$ of $\infty$-stacks
over the site of smooth manifolds (section 4.4 of \cite{dcct}). This contains a canonical line object $\mathbb{A}^1 = \mathbb{R}$,
the \emph{continuum}, abstractly characterized by the fact that the shape modality
exhibits (in the sense of \cite{ShulmanMod}) the corresponding $\mathbb{A}^1$-homotopy localization.
Forming the quotient type by the type of integers yields the \emph{smooth circle group}
$U(1) \simeq \mathbb{R}/\mathbb{Z}$. This being an abelian group type means equivalently that
for all $n \in \mathbb{N}$ there is a pointed $n$-connected type $\mathbf{B}^n U(1)$ such that
$U(1) \simeq \Omega^n \mathbf{B}^n U(1)$ is the $n$-fold loop type.
Write then
$$
  \theta_{\mathbf{B}^n U(1)}
  := \mathrm{fib}(\mathrm{fib}(\epsilon))
  \hspace{2pt}:\hspace{2pt} \mathbf{B}^n U(1) \longrightarrow \flat_{\mathrm{dR}}\mathbf{B}^{n+1} U(1)
$$
for the second homotopy fiber of the co-unit $\epsilon : \flat \mathbf{B}^{n+1}U(1) \longrightarrow \mathbf{B}^{n+1}U(1)$
of the flat co-modality. Cohesion implies that we may think of this as the
\emph{universal Chern-character} for ordinary smooth cohomology (section 3.9.5 in \cite{dcct}). Hence we write
$
  \mathbf{Phases} := \mathbf{B}^n U(1)_{\mathrm{conn}}
$
for the dependent sum of ``all'' homotopy fibers of $\theta_{\mathbf{B}^{n}U(1)}$
(for some choice of ``all'', see section 4.4.16 of \cite{dcct}).
Then a dependent type $\nabla$ over $\mathbf{B}U(1)_{\mathrm{conn}}$ is a \emph{prequantized phase space}
(see section 3.9.13 of \cite{dcct}) in
classical mechanics \cite{Arnold}. An equivalence of dependent types over $\mathbf{B}U(1)_{\mathrm{conn}}$
is a \emph{Hamiltonian symplectomorphism} and a (concrete) function term
$$
  H \hspace{2pt}:\hspace{2pt} \mathbf{B}\mathbb{R} \longrightarrow 
    \underset{\mathbf{B}U(1)_{\mathrm{conn}}}{\prod} \mathbf{B}\mathrm{Equiv}(\nabla,\nabla)
$$
of the function type from the delooping of $\mathbb{R}$ to the delooping of the
dependent product of the type of auto-equivalences of $\nabla$
is equivalently a choice of \emph{Hamiltonian}. It sends the (``time'') parameter
$t : \mathbb{R}$ to the Hamiltonian evolution $\exp(t\{H,-\})$ with Hamilton-Jacobi action
functional $\exp(\tfrac{i}{\hbar} S_t)$ \cite{Arnold}. In the $\infty$-categorical semantics this is given by a
diagram in $\mathbf{H}$ of the following form\footnote{This is a pre-quantization of the \emph{Lagrangian correspondences}
of \cite{Weinstein}.}:
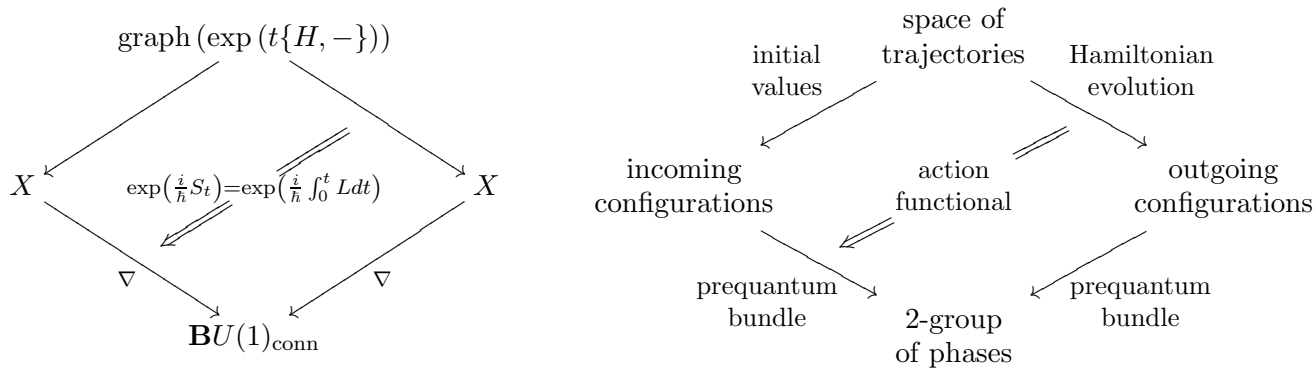
$$
  \hspace{-1cm}
  \raisebox{40pt}{
  \xymatrix{
    & \mathrm{graph}\left(\exp\left(t\{H,-\}\right)\right)
	\ar[dl]
	\ar[dr]_{\ }="s"
	&
	&
    & \mbox{\begin{tabular}{c}space of \\ trajectories\end{tabular}}
	\ar[dl]_{\mbox{\small \begin{tabular}{c}initial \\ values\end{tabular}}}
	\ar[dr]^{\mbox{\small \begin{tabular}{c}Hamiltonian \\ evolution\end{tabular}}}_{\ }="s2"
	\\
	X \ar[dr]_{\nabla}^{\ }="t"
	  &&
	X \ar[dl]^{\nabla}
	&
	\mbox{\begin{tabular}{c}incoming \\ configurations\end{tabular}}
	\ar[dr]_{\mbox{\small \begin{tabular}{c}prequantum \\ bundle \end{tabular}}}^{\ }="t2"
	&&
	\mbox{\begin{tabular}{c}outgoing \\ configurations\end{tabular}}
	\ar[dl]^{\mbox{\small \begin{tabular}{c}prequantum \\ bundle \end{tabular}}}
	\\
	& \mathbf{B}U(1)_{\mathrm{conn}}
	&
	&
    & \mbox{\begin{tabular}{c}2-group \\ of phases\end{tabular}}
	\ar@{=>}|{\exp\left(\frac{i}{\hbar} S_t\right) = \exp\left(\frac{i}{\hbar}\int^t_0 L dt\right)} "s"; "t"
	\ar@{=>}|{\mbox{\small \begin{tabular}{c} action \\ functional \end{tabular}}} "s2"; "t2"
  }
  }
  \,.
$$
Here $X := \underset{\mathbf{B}U(1)_{\mathrm{conn}}}{\sum}\nabla$ is the phase space itself
and $\nabla$ is its \emph{pre-quantum bundle} \cite{hgp}.

This statement concisely captures and unifies a great deal of classical
Hamilton-Lagrange-Jacobi mechanics, as in \cite{Arnold}.
Moreover, when replacing $\mathbf{B}U(1)_{\mathrm{conn}}$ here with $\mathbf{B}^n U(1)_{\mathrm{conn}}$
for general $n \in \mathbb{N}$, then the analogous statement
similarly captures $n$-dimensional classical field theory in its ``covariant'' Hamilton-de Donder-Weyl formulation
on dual jet spaces of the field bundle\footnote{
I am grateful to Igor Khavkine for discussion of this point.} (see e.g. \cite{RomanRoy}).
This is shown in section 1.2.11 of \cite{dcct}.



\end{document}